\documentclass[12pt, a4paper]{article} 
\usepackage{epsf}
\usepackage{cite}
\usepackage{amsmath,amssymb}
\input{colordvi.tex}
\usepackage[usenames,dvipsnames]{color}
\usepackage[dvips]{graphicx}
\usepackage[dvips, bookmarksopen,colorlinks=true,linkcolor=dark_blue,
  citecolor=dark_red,urlcolor=dark_red,linktocpage=false]{hyperref}
\usepackage{comment}
\definecolor{dark_red}{rgb}{0.6,0,0}
\definecolor{dark_blue}{rgb}{0,0,0.6}

\begin{document}

\begin{titlepage}

\begin{center}

\hfill UT-16-36\\
\hfill  IPMU16-0189\\

\vskip .75in

{\Large \bf 
Flaxion: a minimal extension to solve puzzles \\ \vspace{3mm} in the standard model
}

\vskip .75in

{\large Yohei Ema$^{(a)}$, Koichi Hamaguchi$^{(a,b)}$, Takeo Moroi$^{(a,b)}$\\
\vskip 0.1in
and Kazunori Nakayama$^{(a,b)}$}

\vskip 0.25in

$^{(a)}${\em Department of Physics, 
The University of Tokyo,  Tokyo 133-0033, Japan}

\vskip 0.1in

$^{(b)}${\em Kavli Institute for the Physics and Mathematics of the Universe (Kavli IPMU), \\
University of Tokyo, Kashiwa 277--8583, Japan
}


\end{center}
\vskip .5in

\begin{abstract}
  We propose a minimal extension of the standard model which includes
  only one additional complex scalar field, flavon, with
  flavor-dependent global U(1) symmetry.  It not only explains the
  hierarchical flavor structure in the quark and lepton sector
  (including neutrino sector), but also solves the strong CP problem
  by identifying the CP-odd component of the flavon as the QCD axion,
  which we call flaxion.  Furthermore, the flaxion model solves the
  cosmological puzzles in the standard model, i.e., origin of dark
  matter, baryon asymmetry of the universe, and inflation.  We show
  that the radial component of the flavon can play the role of
  inflaton without isocurvature nor domain wall problems.  The dark
  matter abundance can be explained by the flaxion coherent
  oscillation, while the baryon asymmetry of the universe is generated
  through leptogenesis.
\end{abstract}

\end{titlepage}


\renewcommand{\thepage}{\arabic{page}}
\setcounter{page}{1}
\renewcommand{\thefootnote}{$\natural$\arabic{footnote}}
\setcounter{footnote}{0}

\newpage

\section{Introduction} 

There are several puzzles in the standard model (SM) of particle
physics, which may be solved by new physics models based on
(spontaneously broken) symmetries.  Although one may be able to
introduce a new symmetry to solve each puzzle, it is desirable to have
a unified picture of those symmetries from the point of view of
simplicity and minimality, as we suggest in this paper.

One of the mysteries of the SM is the hierarchical flavor structure of
the Yukawa couplings.  The Froggatt-Nielsen mechanism is an attractive
possibility to explain the quark/lepton mass hierarchy and their
mixing matrices~\cite{Froggatt:1978nt}.  It introduces a new complex
scalar field called flavon, whose vacuum expectation value (VEV)
generates the SM Yukawa couplings.  In this model a global Abelian
flavor symmetry U(1)$_F$ is imposed.

Another puzzle in the SM is the strong CP problem in the quantum
chromo dynamics (QCD).  The Peccei-Quinn (PQ)
mechanism~\cite{Peccei:1977hh} utilizes a global U(1) symmetry,
U(1)$_{\rm PQ}$, to solve it; the pseudo Nambu-Goldstone boson
associated with the spontaneous breaking of U(1)$_{\rm PQ}$, called
axion \cite{Weinberg:1977ma, Wilczek:1977pj}, dynamically cancels the
strong CP angle.  Moreover, the PQ model explains the present dark
matter (DM) abundance through the coherent oscillation of the axion
field~\cite{Preskill:1982cy} if the breaking scale of U(1)$_{\rm PQ}$
is at a relevant scale.  It is also remarkable that the right-handed
neutrinos can have large masses through the U(1)$_{\rm PQ}$
breaking~\cite{Langacker:1986rj}.  Thus tiny left-handed neutrino
masses are naturally explained through the seesaw
mechanism~\cite{seesaw}.

In this paper, we propose a new minimal extension of the SM in which
the flavor U(1)$_F$ and the U(1)$_{\rm PQ}$ are unified.  We introduce
only one additional complex scalar field, flavon, charged under the
global U(1)$_F$ whose VEV naturally explains the Yukawa structure.  As
long as this U(1)$_F$ is exact up to the QCD anomaly, its angular
component remains nearly massless, which we call \textit{flaxion}.
Assuming that U(1)$_F$ is anomalous under SU(3)$_C$, the flaxion gets
the potential after the QCD phase transition as ordinary axion and
solves the strong CP problem.  Similar possibility was already pointed
out long ago in Ref.~\cite{Wilczek:1982rv} followed by several
studies~\cite{early_works}.  We use the minimality as our guiding
principle and add only one complex scalar field to the SM.  Then, with
such an additional complex scalar field (as well as right-handed
neutrinos), we show that it is possible to explain the
followings:\footnote
{ A similar approach was made in Refs.~\cite{Salvio:2015cja,Ballesteros:2016euj} in the
  framework of KSVZ axion model, although they did not address the
  flavor structure.  }
(1) Yukawa flavor structure, (2) strong CP problem, (3) neutrino
masses and mixings, (4) dark matter, (5) baryon asymmetry, and (6)
inflation.  In particular, we point out that a successful inflation
takes place by identifying the flavon field as the inflaton.  By
utilizing the idea of attractor inflation \cite{Kallosh:2013hoa}, we
have a phenomenologically viable inflation with successful reheating
consistent with leptogenesis~\cite{Fukugita:1986hr} and without domain
wall nor flaxion isocurvature fluctuation problems.

We emphasize that our model is more economical than other axion
models.  In the KSVZ axion model~\cite{Kim:1979if} heavy vector-like
quarks are necessary, while in the DFSZ axion model~\cite{Dine:1981rt}
we need two Higgs doublets.  In this sense, our model is economical:
addition of only one new scalar field is sufficient to explain the
flavor structure, solve the strong CP problem and provide a good DM
candidate.

This paper is organized as follows.  In Sec.~\ref{sec:flaxion}, we
present our model and derive flavon/flaxion coupling to the SM
particles.  Experimental constraints, in particular flavor-violating
neutral current (FCNC) processes mediated by the flaxion, are also
summarized.  In Sec.~\ref{sec:cosmo}, cosmological aspects of the
flaxion model is discussed.  There we show that the flavon field acts
as the inflaton through the attractor-type mechanism for flattening
the potential without domain wall nor isocurvature problems.  We
conclude in Sec.~\ref{sec:conc} with several remarks.

\section{Flaxion}  \label{sec:flaxion}

\subsection{Model}  \label{sec:model}

The model we consider is described by the following Yukawa terms in the Lagrangian: 
\begin{align}
	-\mathcal L &=  y_{ij}^d\left( \frac{\phi}{M} \right)^{n_{ij}^d} \overline Q_i H d_{Rj}
	+  y_{ij}^u\left( \frac{\phi}{M} \right)^{n_{ij}^u} \overline Q_i \widetilde H u_{Rj} \nonumber\\
	&+ y_{ij}^l\left( \frac{\phi}{M} \right)^{n_{ij}^l} \overline L_i H l_{Rj}
	+  y_{i\alpha}^\nu\left( \frac{\phi}{M} \right)^{n_{i\alpha}^\nu} \overline L_i \widetilde H N_{R\alpha} \nonumber \\
	&+ \frac{1}{2} y_{\alpha\beta}^N\left( \frac{\phi}{M} \right)^{n_{\alpha\beta}^N} M \overline{ N_{R\alpha}^c} N_{R\beta}  +{\rm h.c.}  \, ,
	\label{Lflavon}
\end{align}
where $M$ is a mass scale corresponding to the cut-off scale of this model.
Here $Q_i$, $u_{Ri}$, $d_{Ri}$, $L_i, e_{Ri}$, $N_{R\alpha}$ $(i=1\mathchar`-\mathchar`-3)$ denote
the left-handed quark doublet, right-handed up-type quark,
right-handed down-type quark, left-handed lepton doublet, right-handed
charged lepton and right-handed neutrino, respectively. $H$ denotes
the SM Higgs doublet, and $\widetilde{H}=i\sigma_2 H^*$. Finally,
$\phi$ is a complex scalar field, called flavon, whose VEV
$\left<\phi\right>\equiv v_\phi$ gives rise to the SM Yukawa
couplings~\cite{Froggatt:1978nt}.  The hierarchy of the Yukawa
coupling constants is explained by the smallness of $\epsilon$
defined as
\begin{align}
  \epsilon\equiv \frac{v_\phi}{M}.
\end{align}
We assume all $y_{ij} \sim \mathcal O(1)$ and
$\epsilon \sim 0.2$ to explain the hierarchical structure of the
Yukawa matrix (see App.~\ref{sec:mixing}).  
For a while we do not specify the number of right-handed neutrinos.
The minimal number required to reproduce the experimental results is
two ($\alpha=1,2$)~\cite{Frampton:2002qc,Ibarra:2003up}, while we do
not exclude the possibility of three right-handed neutrinos
($\alpha=1\mathchar`-\mathchar`-3$).  
After $\phi$ and $H$ get VEVs, the mass
matrices are given by
\begin{align}
	m_{ij}^d = y_{ij}^d \epsilon^{n_{ij}^d}v_{\rm EW},~~~m_{ij}^u = y_{ij}^u \epsilon^{n_{ij}^u}v_{\rm EW},~~~
	m_{ij}^l = y_{ij}^l \epsilon^{n_{ij}^l}v_{\rm EW},
\end{align}
where $\left<H\right> \equiv v_{\rm EW} = 174$\,GeV.\footnote{
    The Higgs boson may naturally have mass of $\sim M$ in this
    framework.  The fine-tuning issue to obtain the electroweak scale
    is not addressed in the present study.}

This model possesses a global chiral U(1) symmetry, which we denote by U(1)$_F$, under which the flavon is assumed to have a charge $+1$
and the SM Higgs is neutral.
Denoting the U(1)$_F$ charges of the SM quarks and leptons as $q_{Q_i}, q_{u_i}$ etc., we have the following relations:
\begin{align}
	& n_{ij}^u = q_{Q_i} - q_{u_j}, \\
	& n_{ij}^d = q_{Q_i} - q_{d_j}, \\
	& n_{ij}^l = q_{L_i} - q_{l_j}, \\
	& n_{i\alpha}^\nu = q_{L_i} - q_{N_\alpha} \\
	& n_{\alpha\beta}^N = -q_{N_\alpha} - q_{N_\beta}.
\end{align}
An example for generating the desired quark and lepton masses and the CKM
matrix is\footnote{ The charges of the left-handed fields, $q_{Q_i}$ and
  $q_{L_i}$, are chosen to approximately reproduce the CKM and MNS
  matrices. The other charges, $q_{u_i}, q_{d_i}, q_{l_i}$ are
  determined by $n^f_{ii}\simeq \log(m^f_i/m^t_i)/\log\epsilon$ with
  $\epsilon\simeq 0.23$. (See App.~\ref{sec:mixing}). }
\begin{align}
	\begin{pmatrix}
		q_{Q_1} & q_{Q_2} & q_{Q_3} \\
		q_{u} & q_{c} & q_{t} \\
		q_{d} & q_{s} & q_{b} 
	\end{pmatrix}
	=
	\begin{pmatrix}
		3 & 2 & 0 \\
		-5 & -1 & 0 \\
		-4 & -3 & -3
	\end{pmatrix},  \label{Q_charge}
\end{align}
and
\begin{align}
	\begin{pmatrix}
		q_{L_1} & q_{L_2} & q_{L_3} \\
		q_{e} & q_{\mu} & q_{\tau} 
	\end{pmatrix}
	=
	\begin{pmatrix}
		1 & 0 & 0 \\
		-8 & -5 & -3 
	\end{pmatrix}. \label{L_charge}
\end{align}
That means
\begin{align}
	&n_{ij}^u=\begin{pmatrix}
		8 & 4 & 3 \\
		7 & 3 & 2 \\
		5 & 1 & 0 
	\end{pmatrix},~~
	n_{ij}^d=\begin{pmatrix}
		7 & 6 & 6 \\
		6 & 5 & 5 \\
		4 & 3 & 3 
	\end{pmatrix},~~
	n_{ij}^l=\begin{pmatrix}
		9 & 6 & 4 \\
		8 & 5 & 3 \\
		8 & 5 & 3 
	\end{pmatrix}.
	\label{eq:charge_matrix}
\end{align}
Note that for this charge assignment on the lepton doublets,
the large $\nu_\mu$--$\nu_\tau$ mixing of the neutrino sector is obtained independently of the charges of the right-handed neutrinos~\cite{Yanagida:1998jk,Ramond:1998hs}. (See App.~\ref{sec:neut}).

\subsection{Flavon interactions}

Now let us see the flavon interactions.
Expanding the flavon and Higgs as
\begin{align}
	\phi = v_\phi + \frac{1}{\sqrt{2}}(s+ia),
	\qquad H=\begin{pmatrix}
		0 \\
		v_{\rm EW} + \frac{h}{\sqrt 2}
	\end{pmatrix},
\end{align}
the quark and charged lepton sectors of the Lagrangian (\ref{Lflavon}) are written as
\begin{align}
	-\mathcal L = \sum_{f=u,d,l}
	\left[m_{ij}^f \left(1+\frac{h}{\sqrt{2} v_{\rm EW}}\right)
	+ \frac{m_{ij}^f n_{ij}^f (s+ia)}{\sqrt{2} v_\phi} \right] \overline {f_{Li}} f_{Rj} + {\rm h.c.}
\end{align}
The mass term and Higgs Yukawa interactions are simultaneously diagonalized by the biunitary transformation
\begin{align}
	f_{R_j} \equiv U^f_{ji} f_{R_i}',
	\qquad f_{L_i} \equiv V^f_{ij} f_{L_j}',
	\qquad (V^{f \dagger} m^f U^f)_{ij} = m_i^f \delta_{ij}\,,
\end{align}
but the terms involving the flavon interaction cannot be diagonalized:
\begin{align}
  -\mathcal L &= \sum_{f=u,d,l}
  \left[
    m_{i}^f \left( 1 + \frac{h}{\sqrt{2} v_{\rm EW}} \right) 
    \overline {f'_{Li}} f'_{Ri} 
    +\kappa^f_{ij} \frac{s+ia}{\sqrt{2} v_\phi}\,  
    \overline {f'_{Li}} f'_{Rj}
  \right]
  + {\rm h.c.} \, ,
\end{align}
where the matrix $\kappa^f_{ij}$ is given by
\begin{align}
	\kappa^f_{ij} \equiv V^{f\dagger}_{ik} (m_{kn}^f n_{kn}^f) U^f_{nj}.
\end{align}
Thus the flavon and pseudo-scalar flavon mediate FCNC
processes~\cite{Wilczek:1982rv,ref:flavon_pheno,Feng:1997tn}.  The interaction of the pseudo-scalar
flavon is then written as
\begin{align}
	-\mathcal L = \frac{ia}{\sqrt 2 v_\phi} \sum_{f=u',d',l'}\left[ \left(\kappa^f_{\rm H}\right)_{ij}  \overline f_i \gamma_5 f_j 
	+ \left(\kappa^f_{\rm AH}\right)_{ij} \overline f_i f_j  \right], \label{aff}
\end{align}
where $\kappa^f_{\rm H} = (\kappa^f + \kappa^{f\dagger})/2$ and $\kappa^f_{\rm AH} = (\kappa^f - \kappa^{f\dagger})/2$
are Hermitian and anti-Hermitian parts of $\kappa^f$, respectively. 
Here it may be useful to rewrite the matrices $\kappa^f$ in a simple form.
First note that, the factor $m_{kn}^f n_{kn}^f$ is expressed in a matrix form  as
\begin{align}
m_{kn}^f n_{kn}^f &= \left(\widehat{q}_{Q} m^f - m^f \widehat{q}_f\right)_{kn}\,, \quad f=u,d
\\
m_{kn}^l n_{kn}^l &= \left(\widehat{q}_{L} m^f - m^l \widehat{q}_l\right)_{kn}\,, 
\end{align}
where $(\widehat{q}_X)_{ij}=q_{X_i}\delta_{ij}$ are diagonal matrices.
Then we obtain
\begin{align}
	\kappa^f_{ij} = 
	\left(V^{f\dagger} \widehat{q}_{Q} V^f \right)_{ij} m_j^f 
	- m_i^f \left( U^{f\dagger} \widehat{q}_{f} U^f \right)_{ij}\,,
\end{align}
and
\begin{align}
	&\left(\kappa^f_{\rm H}\right)_{ij} = \frac{1}{2} \left(V^{f\dagger} \widehat{q}_{Q} V^f - U^{f\dagger} \widehat{q}_{f} U^f\right)_{ij} (m_j^f + m_i^f),\\
	&\left(\kappa^f_{\rm AH}\right)_{ij} = \frac{1}{2} \left(V^{f\dagger} \widehat{q}_{Q} V^f + U^{f\dagger} \widehat{q}_{f} U^f\right)_{ij} (m_j^f - m_i^f).
\end{align}
for $f=u,d$.
Expressions for $\kappa^l$ are obtained by replacing $\widehat{q}_Q$ with $\widehat{q}_L$.

\subsection{Flaxion as QCD axion}

The interaction between the pseudo-scalar flavon and quarks (\ref{aff}) yields the effective axion-gluon-gluon interaction
through the triangle anomaly diagram.
The effective interaction is given by
\begin{align}
	\mathcal L 
	= \frac{g_s^2}{32\pi^2} \frac{a}{f_a} G_{\mu\nu}^a \widetilde G^{\mu\nu a},   \label{aGG}
\end{align}
where 
\begin{align}
  f_a \equiv \frac{\sqrt{2}v_\phi}{N_{\rm DW}}
  = \frac{\sqrt{2}\epsilon M}{N_{\rm DW}}\, ,
\end{align}
with the domain-wall number
\begin{align}
  N_{\rm DW} 
	= {\rm Tr}\left(2 \widehat{q}_{Q} - \widehat{q}_{u} - \widehat{q}_{d}\right)
	= {\rm Tr} \left( n^u +  n^d\right)\, ,
\end{align}
which corresponds to the number of the minima of
the potential.  In the model of Sec.~\ref{sec:model}, $N_{\rm
  DW}=26$.  As is well known, after taking the QCD instanton effects
into account, the interaction (\ref{aGG}) results in the axion
potential to cancel the strong CP angle at the potential minimum.
Therefore, we can regard the pseudo-scalar flavon $a$ as the axion
that solves the strong CP problem via the PQ mechanism.  We call $a$
as the flaxion.  The relation between the flaxion mass and the PQ scale is the same as the ordinary QCD axion~\cite{Kim:1986ax}:
\begin{align}
	m_a \simeq 6\times 10^{-6}\,{\rm eV}\left(\frac{10^{12}\,{\rm GeV}}{f_a}\right).
\end{align}
Except that it has a relatively large domain wall
number, its cosmological property is the same as the ordinary
invisible QCD axion.  In particular, the coherent oscillation of the
flaxion can be a good DM candidate.  We will discuss the cosmology of
flaxion and the flavon in Sec.~\ref{sec:cosmo}.

The flaxion-photon coupling is also important for low-energy
phenomenology.  The effective Lagrangian is given by
\begin{align}
	\mathcal L =  g_{a\gamma}\frac{e^2}{32\pi^2} \frac{a}{f_a} F_{\mu\nu}\widetilde F^{\mu\nu},
\end{align}
where~\cite{Kim:1986ax}
\begin{align}
	g_{a\gamma} \equiv \frac{2}{N_{\rm DW}}\sum_{f=u,d,l} \left[ N_f {\rm Tr}\,(n^f) \left(q^{\rm (em)}_f\right)^2\right] - \frac{2(4+z)}{3(1+z)},
\end{align}
with $z\equiv m_u/m_d \simeq 0.56$, $q^{\rm (em)}_f$ the
electromagnetic charge of quarks and leptons, and $N_f=3$ (1) for
quarks (leptons).  For the model presented in Sec.~\ref{sec:model}, we
have $g_{a\gamma} = 113/39-1.95 \simeq 0.95$.  Thus the prospects for
the detection of the flaxion DM are similar to the KSVZ and DFSZ axion model
~\cite{Sikivie:1983ip,Bradley:2003kg,Graham:2013gfa}.

\subsection{Constraints on flaxion}  \label{sec:const}

Phenomenological consequences of the flaxion are similar to the DFSZ
axion, except that the flaxion has FCNC interactions with the quarks and
leptons.  Here we briefly summarize constraints coming from the
flavor-violating process induced by the flaxion and also astrophysical
constraints.

The most stringent bound on $f_a$ may come from the process $K^+\to
\pi^+ a$ mediated by the second term of (\ref{aff}).  In order to
evaluate the matrix element of such a process, we adopt
$\langle\pi(p_\pi) | \overline s \gamma^\mu d (x)| K(p_K)\rangle
\simeq F_1((p_K-p_\pi)^2) e^{-i(p_K-p_\pi)\cdot x} (p_K+p_\pi)^\mu$,
with $F_1(0)\simeq 1$, which holds in the exact SU(3) flavor symmetry
limit.  Then, the matrix element is given by $\mathcal M =
\frac{\left(\kappa^d_{\rm AH}\right)_{12}}{\sqrt{2}v_\phi}\left<
  \pi(p_\pi) | \overline s d | K(p_K)\right> =
\frac{\left(\kappa^d_{\rm
      AH}\right)_{12}}{\sqrt{2}v_\phi}\frac{m_K^2-m_\pi^2}{m_s-m_d}$,
where we have also used the following relation: $\partial_\mu \left<
  \pi(p_\pi) | \overline s \gamma^\mu d | K(p_K)\right> =
(m_s-m_d)\left< \pi(p_\pi) | \overline s d | K(p_K)\right>$.
Consequently, the decay rate is evaluated as
\begin{align}
  \Gamma(K^+\to\pi^+ a) =
  \frac{m_K^3}{32\pi v_\phi^2}
  \left(1-\frac{m_\pi^2}{m_K^2} \right)^3 
  \left|\frac{(\kappa^d_{\rm AH})_{12}}{m_s-m_d} \right|^2,
\end{align}
which gives 
\begin{align}
  {\rm Br}(K^+\to \pi^+ a) \simeq 
  3\times 10^{-10}
  \left( \frac{10^{10}\,{\rm GeV}}{f_a} \right)^2
  \left( \frac{26}{N_{\rm DW}} \right)^2 
  \left|\frac{(\kappa^d_{\rm AH})_{12}}{m_s-m_d} \right|^2.
\end{align}
Comparing with the current experimental bound, Br$(K^+\to \pi^+ a)
\lesssim 7.3\times 10^{-11}$~\cite{Adler:2008zza}, the bound on $f_a$
is given by
\begin{align}
	f_a \gtrsim 2\times 10^{10}\,{\rm GeV} \left( \frac{26}{N_{\rm DW}} \right)\left|\frac{(\kappa^d_{\rm AH})_{12}}{m_s}\right|.
\end{align}
Notice that, because $q_{Q_1}-q_{Q_2}=1$ in order to realize realistic flavor
structure (see App.\ \ref{sec:mixing}), $|(\kappa^d_{\rm
  AH})_{12}/m_s|\sim O(\epsilon)$ (or larger, depending on the
U(1)$_F$ charges of the quarks) assuming no accidental cancellation.
In the near future, it is expected that the NA62
experiment~\cite{Moulson:2016dul} will improve the measurement of $K^+
\to \pi^+ \bar{\nu}\nu$ (and $K^+ \to \pi^+ a$), improving the bound
on $f_a$.

There are also lepton-flavor violating processes mediated by the
flaxion.  Note that processes including double flaxion vertices such
as $\mu-e$ conversion or $\mu\to 3e$ are highly suppressed.  On the
other hand, the decay of muon including the flaxion as a final state
might give a stringent bound.  The three body decay $\mu\to e a
\gamma$~\cite{Goldman:1987hy,Bolton:1988af,Feng:1997tn} might be the
best to constrain the flaxion coupling to the lepton sector.  The
constraint reads Br$(\mu\to ea\gamma) \lesssim 1.1\times 10^{-9}$,
which is translated to~\cite{Goldman:1987hy}
\begin{align}
	f_a \gtrsim 1\times 10^8\,{\rm GeV} \left( \frac{26}{N_{\rm DW}} \right)\left|\frac{(\kappa^l_{\rm AH})_{12}}{m_\mu}\right|.
\end{align}

On the other hand, the observation of SN1987A event at Kamiokande
constrains the flaxion-nucleon coupling, so that the duration of
supernova does not change significantly.  The flaxion-nucleon coupling
is given by
\begin{align}
	\mathcal L = \sum_{N=p,n}\frac{C_N m_N}{f_a} ia \overline N\gamma_5 N,
\end{align}
where
\begin{align}
	&C_p \simeq\left(\frac{ (\kappa^u_{\rm H})_{11} }{m_u N_{\rm DW}}-\frac{1}{1+z} \right)\Delta u + \left(\frac{  (\kappa^d_{\rm H})_{11} }{m_d N_{\rm DW}}-\frac{z}{1+z} \right)\Delta d,\\
	&C_n\simeq\left(\frac{ (\kappa^u_{\rm H})_{11} }{m_u N_{\rm DW}}-\frac{1}{1+z} \right)\Delta d + \left(\frac{ (\kappa^d_{\rm H})_{11} }{m_d N_{\rm DW}}-\frac{z}{1+z} \right)\Delta u,
\end{align}
with $\Delta f$ being the spin content of the nucleon: $S_\mu
\Delta f \equiv \left<N| \bar f\gamma_\mu\gamma_5 f |N\right>$.  They
are given by $\Delta u = 0.85$ and $\Delta d =
-0.41$~\cite{Raffelt:1996wa}, resulting in $C_p\simeq -0.4$ and $|C_n|\ll |C_p|$ for $N_{\rm DW}\gg 1$.  The constraint
reads~\cite{Raffelt:2008}
\begin{align}
  \frac{f_a}{|C_N|} \gtrsim 1\times 10^9\,{\rm GeV},
\end{align}
which is weaker than the constraint from $K^+\to \pi^+ a$.  It is a
striking property of the flaxion, which has flavor-violating
couplings, that the most stringent lower bound on the PQ scale comes
from the flavor physics, not from the SN1987A.
The flaxion-electron coupling is also constrained by the observations of white dwarf stars
so that the cooling of the white dwarf stars due to the flaxion emission does not affect the observed luminosity functions of white dwarf stars too much.
The constraint reads~\cite{Bertolami:2014wua}
\begin{align}
	f_a \gtrsim 7\times 10^7\,{\rm GeV}\left( \frac{26}{N_{\rm DW}} \right)\left|\frac{(\kappa^l_{\rm H})_{11}}{m_e}\right|.
\end{align}
Observations of horizontal branch stars and red-giant stars also put
similar constraints on the flaxion-electron
coupling~\cite{Raffelt:1996wa}.

Let us also comment on the possible constraint from nucleon decay
caused by gauge-invariant baryon- and lepton-number violating higher dimensional operators~\cite{Weinberg:1979sa,Nath:2006ut}. 
If the cutoff scale of these operators are of order $M$, these operators are schematically written as
\begin{align}
	\mathcal L \sim \frac{QQQL}{M^2},~~~\frac{uude}{M^2},~~~\frac{QQue}{M^2},~~~\frac{QLud}{M^2},  \label{p_decay}
\end{align}
which are multiplied by some powers of $\phi/M$ to be consistent with
U(1)$_F$ symmetry.  Due to the suppression factor of powers of
$\epsilon = v_\phi/M$, the effective cutoff scale of these operators
can be much higher than $M$.  For the charge assignments of
(\ref{Q_charge}) and (\ref{L_charge}), the most dangerous operator is
the last one in (\ref{p_decay}), which is suppressed only by
$\epsilon^5$ for the first generation quarks and leptons.  Therefore,
the effective cutoff scale of this operator is $M_{\rm eff}\sim
\epsilon^{-2.5} M \sim 40\times M$ and hence we need $M \gtrsim
5\times 10^{14}$\,GeV to avoid the too rapid proton
decay~\cite{Takhistov:2016eqm}.  This is roughly consistent with the
phenomenologically preferred value $M\sim
10^{14}\text{--}10^{17}$\,GeV, as shown in the next section.  One
should also note that this suppression factor crucially depends on the
U(1)$_F$ charge assignments on the quarks and leptons.  As shown in
App.~\ref{sec:CKM}, we have a freedom of constant shift of all the
$(q_{Q_i}, q_{u_i}, q_{d_i})$ without affecting $n^u_{ij}$ and
$n^d_{ij}$.  Using this freedom, it is possible to suppress all of the
operators in (\ref{p_decay}) further.  Since the Lagrangian~(\ref{Lflavon}) 
depends only on the combination $n_{ij}$, all the
phenomenological constraints discussed so far, except for the nucleon
decay, remain intact with such a shift of U(1)$_F$ charges.

\section{Flaxion and flavon cosmology}  \label{sec:cosmo}

\subsection{Flaxion as dark matter}

Let us discuss cosmological consequences of the present
model~\cite{Kawasaki:2013ae}.  As in the case of ordinary QCD axion,
the flaxion starts to oscillate around the minimum of the potential.
Its present density is given by~\cite{Turner:1985si}
\begin{align}
  \Omega_a h^2 = 0.18\,
  \theta_i^2 \left( \frac{f_a}{10^{12}\,{\rm GeV}}\right)^{1.19},
\label{eq:Omega}
\end{align}
where $\theta_i$ denotes the initial misalignment angle which takes
the value $0\leq\theta_i <2\pi$.  Thus, the flaxion oscillation can be
dark matter for 
$f_a\sim O(10^{12}\text{--}10^{15})$\,GeV, assuming $\theta_i\simeq
O(0.01$--$1)$.

As discussed in the previous section, the decay constant of the
flaxion is related to the parameters in the flavon potential.  For
$N_{\rm DW}=26$ and $\epsilon\sim 0.2$, for example, the flaxion dark
matter is realized when 
$v_\phi\sim O(10^{13}\text{--}10^{16})\ {\rm GeV}$ 
and 
$M\sim O(10^{14}\text{--}10^{17})\ {\rm GeV}$.


\subsection{Isocurvature and domain wall problem}

Since the domain wall number is larger than unity, one may require
that the U(1)$_F$ symmetry be spontaneously broken during inflation to
avoid the serious domain wall problem.  In this case there is a
stringent constraint on the inflation energy scale so that the flaxion
does not acquire too large isocurvature fluctuations.  The recent
constraint from the Planck result reads $\sqrt{\mathcal P_{S}/\mathcal
  P_\zeta} \lesssim 0.18$ with $\mathcal P_{\zeta} \simeq 2.2\times
10^{-9}$, where $\mathcal P_{S}$ and $\mathcal P_{\zeta}$ are the
dimensionless power spectrum of the (uncorrelated) DM isocurvature and
curvature perturbations, respectively~\cite{Ade:2015lrj}.  If the
flavon field settles down to the potential minimum during inflation,
we have
\begin{align}
	\mathcal P_S \simeq  \left( \frac{H_{\rm inf}}{\pi f_a \theta_i} \right)^2
	\left(\frac{\Omega_a  h^2}{\Omega_{\rm CDM} h^2}\right)^2,
\end{align}
where $\Omega_a  h^2$ is given by Eq.~\eqref{eq:Omega} and $\Omega_{\rm CDM} h^2\simeq 0.12$.
Thus the inflationary scale is bounded as
\begin{align}
	H_{\rm {inf}} \lesssim 3\times 10^7\,{\rm GeV}\,\theta_i^{-1} \left( \frac{10^{12}\,{\rm GeV}}{f_a} \right)^{0.19}.
\end{align}
Notice that this constraint is based on the assumption that the flavon
already settles down to its potential minimum during inflation.
However, the dynamics of the flavon field can be non-trivial during
and after inflation and it can significantly modify the constraint.  Below
we see that the flavon itself can play the role of inflaton, avoiding
this isocurvature bound.

\subsection{Flavon inflation}

So far we have assumed that the inflaton sector is independent of the SM + flavon sector.
More interestingly, it may be possible to identify the flavon itself as the inflaton.\footnote{
	A flavon inflation was considered in Ref.~\cite{Antusch:2008gw} in a different context.
}
First of all, one should note that large field inflation in which $\varphi\equiv \sqrt{2}{\rm Re(\phi)}$ rolls down from $\varphi \gg v_\phi$
would be dangerous, since during the reheating stage the flavon passes through the origin $\varphi=0$ many times
and it leads to the nonthermal symmetry restoration through the parametric resonant enhancement of the flaxion field~\cite{Kofman:1995fi,Kasuya:1996ns}.
Thus the domain wall problem arises after the QCD phase transition in such a case.
On the other hand, the small-field inflation in which $\varphi$ rolls down from near the origin toward the potential minimum may be possible.
Although there is no domain wall problem in this case, the flavon self coupling constant needs to be very small
and also the flaxion isocurvature perturbation tends to be too large because it is enhanced due to the smallness of $\varphi$ during inflation.

Here we propose a nonminimal large-field inflation model which avoids
these difficulties.  The idea is to extend the flavon kinetic term to
effectively flatten the potential at large field
value~\cite{Takahashi:2010ky}.  In an extreme case in which the
kinetic term has a pole at some field value, the effective potential
becomes completely flat around the pole after the canonical
normalization, and it leads to a class of large field inflation with
best-fit value of the scalar spectral index~\cite{Kallosh:2013hoa}.

\begin{figure}
\begin{center}
\includegraphics[scale=0.7]{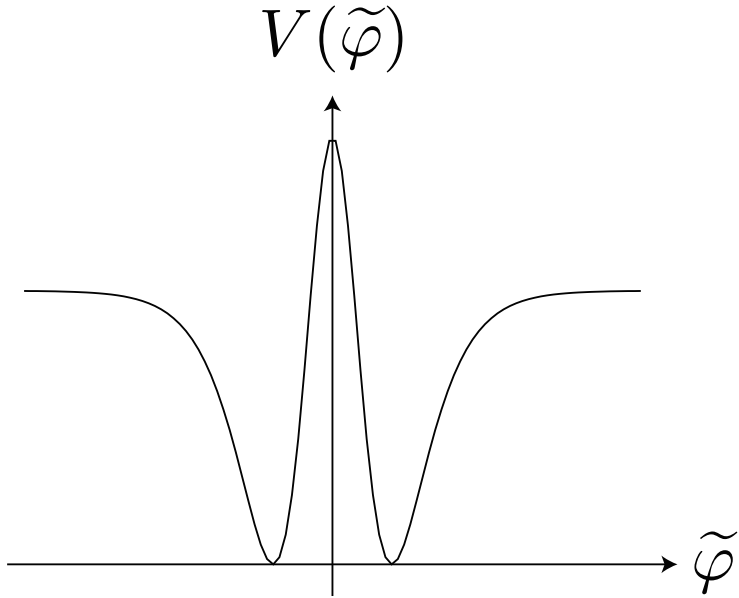}
\end{center}
\caption {
	Schematic picture of the flavon potential for successful inflation.
}
\label{fig:pot}
\end{figure}

Here, we adopt the following Lagrangian:
\begin{align}
	\mathcal L = -\frac{|\partial\phi|^2}{\left(1-\frac{|\phi|^2}{\Lambda^2}\right)^2} - \lambda_\phi \left( |\phi|^2-v_\phi^2 \right)^2.
	\label{eq:L_inflaton}
\end{align}
After the canonical normalization, the flavon potential may be rewritten as
\begin{align}
	\mathcal L = -\frac{(\partial\widetilde\varphi)^2}{2} - \lambda_\phi\left[ \Lambda^2 \tanh^2\left(\frac{\widetilde\varphi}{\sqrt 2\Lambda}\right)-v_\phi^2 \right]^2,
\end{align}
where
\begin{align}
	\frac{\varphi}{\sqrt{2}\Lambda} \equiv \tanh \left(\frac{\widetilde\varphi}{\sqrt{2}\Lambda}\right).
\end{align}
Thus the potential is flat for $\widetilde\varphi \gg \Lambda$ and inflation can take place there. 
If $v_\phi < \Lambda < \sqrt{2}v_\phi$, the potential height at large field limit is lower than that at the origin (see Fig.~\ref{fig:pot}),
hence the flavon does not pass through the origin after inflation.
Thus there is no domain wall problem in this case.
Note that the potential minimum in terms of $\widetilde\varphi$ is 
\begin{align}
	\frac{\left< \widetilde\varphi\right>}{\sqrt 2} = \Lambda\tanh^{-1}\left(\frac{v_\phi}{\Lambda}\right).
\end{align}
As we will discuss in the following, $\Lambda$ is found to be of the
same order of $v_\phi$ in the parameter region of our interest, and
hence $\left< \widetilde\varphi\right>\sim O(v_\phi)$.

We also note here that, due to the structure of the kinetic term in
the present model, the field $a$ is not canonically normalized.  The
canonically-normalzed flaxion field around the vacuum is $\tilde a
\equiv a/\Delta$, where
\begin{align}
  \Delta\equiv 1-v_\phi^2/\Lambda^2.
\end{align}
Thus, the flaxion interactions (as well as the decay rate) given in
the previous section should take account of the correction factor
$\Delta$.  For the case of our interest, however, $\Delta\sim O(1)$
and hence the discussion given in the previous section is
qualitatively unchanged.

We can analyze the slow-roll inflation dynamics as usual~\cite{Liddle:2000cg}. The flavon field value during inflation is calculated as
\begin{align}
	\widetilde\varphi_N \simeq \frac{\Lambda}{\sqrt{2}}\ln \left( \frac{16N_e M_P^2}{\Lambda^2-v_\phi^2} \right),
\end{align}
where $N_e \sim 50$ -- $60$ denotes the $e$-folding number at which the
present horizon scale exits the horizon.  The scalar spectral index
$n_s$ and the tensor-to-scalar ratio $r$ are given by
\begin{align}
	n_s \simeq 1 - \frac{2}{N_e},~~~~~~~~~~r\simeq \frac{4}{N_e^2}\left( \frac{\Lambda}{M_P} \right)^2.
\end{align}
Thus the scalar spectral index falls into the Planck best-fit region
while the tensor-to-scalar ratio is too small to be detected.  The
dimensionless power spectrum of the curvature perturbation is given by
\begin{equation}
	\mathcal P_\zeta \simeq \frac{N_e^2}{6\pi^2}\frac{\lambda_\phi(\Lambda^2-v_\phi^2)^2}{\Lambda^2 M_P^2}.
\end{equation}
In order to reproduce the observed magnitude of the curvature
perturbation, $\mathcal P_{\zeta} \simeq 2.2\times 10^{-9}$~\cite{Ade:2015lrj},
\begin{align}
	\lambda_\phi \simeq 3\times 10^{-2}\left( \frac{50}{N_e} \right)^2\left( \frac{10^{14}\,{\rm GeV}}{\Lambda} \right)^2
	\left( \frac{\Lambda^2}{\Lambda^2-v_\phi^2} \right)^2.
\end{align}
In order for $\lambda_\phi$ to be in the perturbative range, we must have $\Lambda \gtrsim 10^{13}\,$GeV,
meaning $f_a \sim \Lambda/N_{\rm DW} \gtrsim 5\times 10^{11}$\,GeV.
This is consistent with the scale inferred from the flaxion DM density~\eqref{eq:Omega}.
The inflation scale is given by
\begin{align}
	H_{\rm inf} \simeq 5\times 10^8\,{\rm GeV}\left( \frac{\Lambda}{10^{14}\,{\rm GeV}} \right).
\end{align}

\subsection{Suppression of isocurvature perturbation}

Here we show that the isocurvature perturbation of the flaxion is
highly suppressed due to the peculiar structure of the kinetic term.
We parametrize the complex flavon field as $\phi = \varphi
e^{i\Theta}/\sqrt{2}$. Then, the action for the phase $\Theta$ is
given by
\begin{align}
  \mathcal{L} = \frac{\Lambda^2}{4}\sinh^2
  \left(\frac{\sqrt{2}\widetilde\varphi}{\Lambda}\right)
  \left(\partial \Theta\right)^2.
	\label{eq:phase_Lag}
\end{align}
Since $\widetilde\varphi$ slow-rolls during inflation, we may regard
the prefactor in~\eqref{eq:phase_Lag} as a constant.  Then, the
canonically normalized field during inflation is given by
\begin{align}
  \tilde{a}_{\rm inf} = 
  \frac{\Lambda}{\sqrt{2}}
  \sinh\left(\frac{\sqrt{2}\widetilde\varphi}{\Lambda}\right)\Theta.
\end{align}
The canonical field $\tilde{a}_{\rm inf}$ acquires long-wavelength
fluctuations of $H_\mathrm{inf}/2\pi$ during inflation, and hence the
original phase $\Theta$ fluctuates as
\begin{align}
  \mathcal{P}_{\delta\Theta}
  &\simeq \frac{2H_\mathrm{inf}^2}{\pi^2\Lambda^2}
  \exp\left(-\frac{2\sqrt{2}\widetilde\varphi_N}{\Lambda}\right) 
  \simeq \frac{H_\mathrm{inf}^2}{128\pi^2\Lambda^2}
  \left(\frac{\Lambda^2 - v_\phi^2}{N_eM_P^2}\right)^2,
\end{align}
where $\mathcal{P}_{\delta\Theta}$ is the power spectrum of
$\delta\Theta$.  (Here, we have used $\widetilde\varphi \gg \Lambda$
during inflation.)  Since it is related to the fluctuation of the
initial misalignment angle as $\delta \theta_i =
N_\mathrm{DW}\delta\Theta$,
 the ratio of the DM isocurvature
perturbation to the curvature perturbation is estimated as
\begin{align}
	\frac{\mathcal{P}_S}{\mathcal{P}_\zeta}
	\simeq \frac{R_a^2N_\mathrm{DW}^2}{64\theta_i^2N_e^4} \frac{\left(\Lambda^2 - v_\phi^2\right)^2}{M_P^4},
\end{align}
which is highly suppressed.\footnote{
In this scenario, the spectrum of the flaxion isocurvature fluctuation is blue. 
However, even at the smallest scale the isocurvature perturbation is small enough.
} 
Thus, the observational bound is safely
satisfied and the flaxion can be the dominant component of DM.

\subsection{Reheating after flavon inflation}

Finally let us discuss the reheating after flavon inflation.
There are mainly three decay modes of the flavon: decay into right-handed neutrinos, decay into Higgs bosons and decay into flaxions.
Other decay modes are suppressed either by the loop factor or the final state fermion masses.

The flavon partial decay rate into the right-handed neutrino pair is given by 
\begin{align}
  \Gamma(\widetilde\varphi\to N_{R} N_{R}) \simeq
  \sum_{\alpha\beta}
  \frac{|y^N_{\alpha\beta} n^N_{\alpha\beta}\epsilon^{n^N_{\alpha\beta}-1}|^2}
  {32\pi}\Delta^2 m_\varphi,
\end{align}
where the flavon mass around the potential minimum is given by
\begin{align}
	m_\varphi^2 = 4\lambda_\phi v_\phi^2 \Delta^2.
\end{align}
Note that $m_\varphi \sim 3\times 10^{13}\,{\rm GeV} (v_\phi/\Lambda)$ is almost independent of the overall scale $\Lambda$.
Here we have assumed that the flavon is heavier than the right-handed neutrino: 
$4\lambda_\phi \Delta^2 \gtrsim (y^N_{\alpha\alpha}\epsilon^{n^N_{\alpha\alpha}-1})^2$.
The partial decay rate of flavon into the Higgs bosons depend on the additional potential term\footnote{
	This term potentially leads to the vacuum decay through the resonant enhancement of the Higgs fluctuation~\cite{Herranen:2015ima,Ema:2016kpf,Kohri:2016wof,Enqvist:2016mqj}.
	However, the same term along with large VEV of $\phi$ can ensure the absolute stability of the Higgs potential~\cite{Lebedev:2012zw}.
	Note also that there must be a large bare mass term of the Higgs to cancel the flavon-induced mass term so that it obtains
	the electroweak scale VEV.
}
\begin{align}
	V = \lambda_{\phi H} |\phi|^2 |H|^2.
\end{align}
We find the partial decay rate into the Higgs boson pair as
\begin{align}
  \Gamma(\widetilde\varphi\to HH) \simeq 
  \frac{\Delta^2}{8\pi}\frac{\lambda_{\phi H}^2 v_\phi^2}{m_\varphi} 
  \simeq \frac{1}{32\pi} \frac{\lambda_{\phi H}^2}{\lambda_\phi} m_\varphi,
\end{align}
where we have taken account of the four real degrees of freedom in the SM Higgs doublet.\footnote{
	This coupling radiatively affects the flavon potential.
	If it is substantially large and $\lambda_\phi$ is too small, the flavon potential can be dominated by the
	radiatively-induced effective potential.
}
On the other hand, the flavon partial decay rate into the flaxion pair is given by
\begin{align}
	\Gamma(\widetilde\varphi\to aa) \simeq \frac{\Delta^2}{32\pi} \frac{m_\varphi^3}{v_\phi^2} \simeq \frac{\lambda_\phi}{8\pi} \Delta^4 m_\varphi. 
\end{align}
Thus the total decay width of the flavon is
\begin{align}
	\Gamma_{\widetilde\varphi} \simeq \left(
	\sum_{\alpha,\beta}
	 \left|
	 y^N_{\alpha\beta} n^N_{\alpha\beta}\epsilon^{n^N_{\alpha\beta}-1}\right|^2
	 \frac{\Delta^2}{4} + \frac{\lambda_{\phi H}^2}{4\lambda_\phi} + \lambda_\phi \Delta^4\right)
	\frac{m_\varphi}{8\pi}.
\end{align}
This is typically much larger than $H_{\rm inf}$ and hence the reheating is completed almost instantaneously after inflation.
Thus the reheating temperature, $T_{\rm R}$, can be as high as $10^{12}\text{--}10^{14}$\,GeV in our scenario.
Flaxions are thermalized through interactions with Higgs and right-handed neutrinos and there is no problem of flaxion dark radiation overproduction.



Lastly let us discuss thermal leptogenesis in the present scenario.
The final baryon asymmetry through the leptogenesis from the decay of right-handed neutrinos is given by~\cite{Buchmuller:2004nz}
\begin{align}
	\frac{n_B}{s}\simeq \epsilon_1 \kappa_f \frac{28}{79}\left(\frac{n_{N_1}}{s}\right)_{\rm th}\simeq 1.3\times 10^{-3} \epsilon_1 \kappa_f,
\end{align}
where $(n_{N_1}/s)_{\rm th}$ is the abundance of the right-handed neutrino in thermal equilibrium,
$\epsilon_1$ denotes the lepton asymmetry generated by per right-handed neutrino decay and $\kappa_f$ denotes the efficiency factor.
The asymmetry parameter is calculated as
\begin{align}
	\epsilon_1 
	= \frac{3}{16\pi} \frac{m_{N_1} m_{\nu_3}}{v_{\rm EW}^2}\delta_{\rm eff}
	\simeq 1\times 10^{-4}\left( \frac{m_{N_1}}{10^{12}\,{\rm GeV}} \right)\left( \frac{m_{\nu 3}}{0.05\,{\rm eV}} \right)\delta_{\rm eff},
\end{align}
where $\delta_{\rm eff}$ is the effective CP angle which satisfies $\delta_{\rm eff} \le 1$ for $m_{N_1}\ll m_{N_{2(3)}}$~\cite{Hamaguchi:2001gw,Davidson:2002qv}.\footnote{
If the mass of $N_1$ is degenerated with $N_2$, the asymmetry is enhanced~\cite{Covi:1996fm}.
This can happen in our case if U(1)$_F$ charges of right-handed neutrinos are the same.
}
On the other hand, the efficiency factor $\kappa_f$ crucially depends on the effective neutrino mass
$\widetilde m_{\nu 1} \equiv \sum_k |\epsilon^{n^\nu_{k1}}y_{k1}^\nu|^2 v_{\rm EW}^2/m_{N_1}$.
In the present scenario, it is roughly given by 
$\widetilde m_{\nu 1}\sim \sum_k \epsilon^{2q_{L_k}} v_{\rm EW}^2/M\sim m_{\nu_3}$. 
(See App.~\ref{sec:mixing}.) This corresponds to a so-called strong washout regime ($\widetilde m_{\nu 1}\gtrsim
m_*\simeq 1\times 10^{-3}~{\rm eV}$), 
where the efficiency factor is approximately given by $\kappa_f \sim 0.02\times  (\widetilde m_{\nu 1}/0.01~{\rm eV})^{-1.1}$~\cite{Buchmuller:2004nz}. For $\widetilde m_{\nu 1}\sim m_{\nu_3}\sim 0.05~{\rm eV}$, we obtain $\kappa_f\sim 3\times 10^{-3}$.
Therefore,  the observed baryon asymmetry $n_B/s\simeq 9\times 10^{-11}$ can be obtained 
for $m_{N_1}\sim O(10^{12})$\,GeV.\footnote{For this mass scale, none of the charged lepton 
Yukawa coupling is in equilibrium, and the flavor effect~\cite{ref:flavor} can be neglected.}
This can be obtained, for instance, by taking $q_{N_1}=1-5$ for $M\sim O(10^{14}\text{--}10^{17})\,{\rm GeV}$.

\section{Conclusions and discussion}  \label{sec:conc}

We have shown that a simple QCD axion model in which U(1)$_{\rm PQ}$
is identified with Abelian flavor symmetry U(1)$_F$ solves and
explains puzzles in the SM.  The model contains only one additional
complex scalar and right-handed neutrinos.  Inflation can successfully
happen without domain wall nor isocurvature problems.

Here are some remarks.  In this paper, we assume that there is only
one Higgs doublet. Although this is a minimal choice, if there are
additional Higgs doublets, we can assign the $\mathrm{U}(1)_F$ charges
to the Higgses so that $N_\mathrm{DW} = 1$.  As an example, we
consider a two Higgs doublet model (2HDM)~\cite{Branco:2011iw} with
the following Yukawa interactions (the so-called type-I\hspace{-.1em}I
or type-Y 2HDM):
\begin{align}
	-\mathcal{L} = y_{ij}^d\left( \frac{\phi}{M} \right)^{n_{ij}^d} \overline Q_i H_d d_{Rj}
	+  y_{ij}^u\left( \frac{\phi}{M} \right)^{n_{ij}^u} \overline Q_i H_u u_{Rj}.  \label{2HDM}
\end{align}
If we assign the $\mathrm{U}(1)_F$ charges $q_{H_u}$ and $q_{H_d}$ on $H_u$ and $H_d$ respectively,
we obtain
\begin{align}
	n_{ij}^d &= q_{Q_i} - q_{d_j} - q_{H_d}, \\
	n_{ij}^u &= q_{Q_i} - q_{u_j} - q_{H_u}.
\end{align}
We may keep $n_{ij}^f$ the same as those in~\eqref{eq:charge_matrix} by shifting
the charges of the right-handed quarks as $q_{f_i} \rightarrow q_{f_i} - q_{H_f}$.\footnote{
If $\tan \beta \equiv \langle H_u\rangle/\langle H_d\rangle$ is not of order unity,
the ratio of the overall normalization of $n_{ij}^d$ and $n_{ij}^u$ can be much different 
than~\eqref{eq:charge_matrix}.
}
Then, the domain wall number is given by
\begin{align}
	N_\mathrm{DW} 
	= \left\lvert {\rm Tr}\left(2 \widehat{q}_{Q} - \widehat{q}_{u} - \widehat{q}_{d}\right)\right\rvert
	= \left\lvert 26 + 3\left(q_{H_u} + q_{H_d}\right)\right\rvert.
\end{align}
Thus, the domain wall number is $N_\mathrm{DW} = 1$ if we take $q_{H_u} + q_{H_d} = -9$.
In this case, there is no cosmological domain wall problem even if the PQ symmetry is restored during inflation,
as long as $f_a < (4.6-7.2)\times 10^{10}\,$GeV~\cite{Kawasaki:2014sqa}
and hence there can be a variety of cosmological scenarios.

Although the minimality is lost, it is also easy to embed the theory into the supersymmetry (SUSY) framework.
We can just interpret the Lagrangian (\ref{2HDM}) as the superpotential written by the chiral superfields.
Since there are two Higgs doublets in minimal SUSY SM, we can choose the U(1)$_F$ charges so that $N_\mathrm{DW} = 1$ as just shown above.
In this case, the $\mu$-term can be generated by the superpotetnial $W \sim (\phi/M)^9 M H_u H_d$,
which may be compatible with high-scale SUSY scenario in which the soft mass scale is $O(100-1000)\,$TeV.
The potential of the flavon can be generated by introducing 
$\bar\phi$ and also a ``stabilizer field'' $X$, 
which have U(1)$_F$ charges $-1$ and $0$ respectively, and assume the superpotential
\begin{align}
	W = \lambda X (\phi\bar\phi - v_\phi^2).
\end{align}
After they get soft SUSY breaking masses, they are stabilized at $\phi
\sim \bar\phi \sim v_\phi$.  Since $\bar\phi$ is oppositely charged
under U(1)$_F$, it cannot directly couple to SM Yukawa terms.  Note
that, with the present assignments of U(1)$_F$ charges, off-diagonal
elements of the squark mass matrix are not suppressed enough to avoid
SUSY flavor problem if the mass scale of the SUSY particiles is around
TeV.  Such a problem can be solved by high-scale SUSY or flavor-blind
mediation model (like gauge mediation).  (Otherwise one may adopt a
different flavor symmetry to suppress the off-diagonal elements of the
sfermion mass matrix.)  Cosmology of this class of models will be
non-trivial due to the presence of sflaxion and flaxino which appear
in the flavon supermultiplet, although the detailed investigation is
beyond the scope of this paper.

\section*{Acknowledgments}
We thank Natsumi Nagata for helpful discussion.
This work was supported by the Grant-in-Aid for Scientific Research on Scientific
Research A (No.26247038 [KH], No.26247042 [KN], No.16H02189 [KH]),
Scientific Research C (No.26400239 [TM]), Young Scientists B
(No.26800121 [KN], No.26800123 [KH]) and Innovative Areas (No.26104001
[KH], No.26104009 [KH and KN], No.15H05888 [KN], No.16H06490 [TM]), and
by World Premier International Research Center Initiative (WPI
Initiative), MEXT, Japan.
The work of YE was supported in part by JSPS Research Fellowships
for Young Scientists and by the Program for Leading Graduate Schools, MEXT, Japan.  

\appendix

\section{Quark and lepton masses and mixings}  \label{sec:mixing}

\subsection{Quark and charged lepton masses and CKM matrix}  \label{sec:CKM}

For general U(1)$_F$ charge assignments on the quark fields
\begin{align}
	\begin{pmatrix}
		q_{Q_1} & q_{Q_2} & q_{Q_3} \\
		q_{u} & q_{c} & q_{t} \\
		q_{d} & q_{s} & q_{b} 
	\end{pmatrix},
\end{align}
with assumption $q_{Q_i} \geq q_{Q_j} \geq 0$ and $q_{f_i} \leq q_{f_j}\leq 0$ for $i < j$,
the quark mass matrix (normalized by $v_{\rm EW}$) is expressed and decomposed as
\begin{align}
	m_{ij}^d &\sim \epsilon^{n_{ij}^d} \sim V^d {\rm diag}(m_d) U^{d\dagger} 
        \nonumber \\
	&\sim \begin{pmatrix}
		1 &\epsilon^{q_{Q_1}-q_{Q_2}} & \epsilon^{q_{Q_1}-q_{Q_3}} \\
		\epsilon^{q_{Q_1}-q_{Q_2}} & 1 & \epsilon^{q_{Q_2}-q_{Q_3}} \\
		\epsilon^{q_{Q_1}-q_{Q_3}} & \epsilon^{q_{Q_2}-q_{Q_3}} & 1 
	\end{pmatrix}
	\begin{pmatrix}
		 \epsilon^{q_{Q_1}-q_d} & 0 & 0\\
		0 &  \epsilon^{q_{Q_2}-q_s} & 0\\
		0 & 0 & \epsilon^{q_{Q_3}-q_b}
	\end{pmatrix}
	\begin{pmatrix}
		1 &\epsilon^{q_{s}-q_{d}} & \epsilon^{q_{b}-q_{d}} \\
		\epsilon^{q_{s}-q_{d}} & 1 & \epsilon^{q_{b}-q_{s}} \\
		\epsilon^{q_{b}-q_{d}} & \epsilon^{q_{b}-q_{s}} & 1 
	\end{pmatrix},
\end{align}
\begin{align}
	m_{ij}^u &\sim \epsilon^{n_{ij}^u} \sim V^u {\rm diag}(m_u) U^{u\dagger}
        \nonumber \\
	&\sim \begin{pmatrix}
		1 &\epsilon^{q_{Q_1}-q_{Q_2}} & \epsilon^{q_{Q_1}-q_{Q_3}} \\
		\epsilon^{q_{Q_1}-q_{Q_2}} & 1 & \epsilon^{q_{Q_2}-q_{Q_3}} \\
		\epsilon^{q_{Q_1}-q_{Q_3}} & \epsilon^{q_{Q_2}-q_{Q_3}} & 1 
	\end{pmatrix}
	\begin{pmatrix}
		 \epsilon^{q_{Q_1}-q_u} & 0 & 0\\
		0 &  \epsilon^{q_{Q_2}-q_c} & 0\\
		0 & 0 & \epsilon^{q_{Q_3}-q_t}
	\end{pmatrix}
	\begin{pmatrix}
		1 &\epsilon^{q_{c}-q_{u}} & \epsilon^{q_{t}-q_{u}} \\
		\epsilon^{q_{c}-q_{u}} & 1 & \epsilon^{q_{t}-q_{c}} \\
		\epsilon^{q_{t}-q_{u}} & \epsilon^{q_{t}-q_{c}} & 1 
	\end{pmatrix}.
\end{align}
Thus the CKM matrix is given by
\begin{align}
	V_{\rm CKM} = V^{u\dagger} V^d \sim
	\begin{pmatrix}
		1 &\epsilon^{q_{Q_1}-q_{Q_2}} & \epsilon^{q_{Q_1}-q_{Q_3}} \\
		\epsilon^{q_{Q_1}-q_{Q_2}} & 1 & \epsilon^{q_{Q_2}-q_{Q_3}} \\
		\epsilon^{q_{Q_1}-q_{Q_3}} & \epsilon^{q_{Q_2}-q_{Q_3}} & 1 
	\end{pmatrix},
\end{align}
which depends only on the charges of the left-handed quarks.
Taking account of $O(1)$ Yukawa couplings, it well reproduces observed values of the CKM matrix elements for
\begin{align}
	\begin{pmatrix}
		q_{Q_1} & q_{Q_2} & q_{Q_3}
	\end{pmatrix}
	=
	\begin{pmatrix}
		q_{Q_3}+3 & q_{Q_3}+2 & q_{Q_3}
	\end{pmatrix},
\end{align}
and $\epsilon \simeq 0.23$.
Charges of right-handed quarks are chosen so that the quark mass eigenvalues are consistent with observed values:
\begin{align}
	q_{Q_1}-q_d=7,~~~q_{Q_2}-q_s=5,~~~q_{Q_3}-q_b=3,\\
	q_{Q_1}-q_u=8,~~~q_{Q_2}-q_c=3,~~~q_{Q_3}-q_t=0.
\end{align}
Still we have a degree of freedom to choose $q_{Q_3}$, corresponding to the overall constant shift of $(q_{Q_i}, q_{u_i},q_{d_i})$.\footnote{
	In other words, we can arbitrarily add baryon charges to the U(1)$_F$ charges.
}
A particular example with $q_{Q_3}=0$ is given in (\ref{Q_charge}).

Similarly, for general U(1)$_F$ charge assignments on the leptons
\begin{align}
	\begin{pmatrix}
		q_{L_1} & q_{L_2} & q_{L_3} \\
		q_{e} & q_{\mu} & q_{\tau}
	\end{pmatrix},
\end{align}
with assumption $q_{L_i} \geq q_{L_j} \geq 0$ and $q_{f_i} \leq q_{f_j}\leq 0$ for $i < j$,
the charged lepton mass matrix (normalized by $v_{\rm EW}$) is decomposed as
\begin{align}
	m_{ij}^l &\sim \epsilon^{n_{ij}^l} \sim V^l {\rm diag}(m_l) U^{l\dagger}
        \nonumber \\
	&\sim \begin{pmatrix}
		1 &\epsilon^{q_{L_1}-q_{L_2}} & \epsilon^{q_{L_1}-q_{L_3}} \\
		\epsilon^{q_{L_1}-q_{L_2}} & 1 & \epsilon^{q_{L_2}-q_{L_3}} \\
		\epsilon^{q_{L_1}-q_{L_3}} & \epsilon^{q_{L_2}-q_{L_3}} & 1 
	\end{pmatrix}
	\begin{pmatrix}
		 \epsilon^{q_{L_1}-q_e} & 0 & 0\\
		0 &  \epsilon^{q_{L_2}-q_\mu} & 0\\
		0 & 0 & \epsilon^{q_{L_3}-q_\tau}
	\end{pmatrix}
	\begin{pmatrix}
		1 &\epsilon^{q_{\mu}-q_{e}} & \epsilon^{q_{\tau}-q_{e}} \\
		\epsilon^{q_{\mu}-q_{e}} & 1 & \epsilon^{q_{\tau}-q_{\mu}} \\
		\epsilon^{q_{\tau}-q_{e}} & \epsilon^{q_{\tau}-q_{\mu}} & 1 
	\end{pmatrix}.
\end{align}
The observed charged lepton masses are reproduced for
\begin{align}
	\begin{pmatrix}
		q_{L_1} & q_{L_2} & q_{L_3}
	\end{pmatrix}
	=
	\begin{pmatrix}
		q_{e}+9 & q_{\mu}+5 & q_{\tau}+3
	\end{pmatrix}.
\end{align}
The charges of left-handed leptons are partly constrained from the neutrino mass matrix, as shown below.

\subsection{Neutrino masses and mixing}
\label{sec:neut}

First let us consider the minimal case of two right-handed neutrinos: $N_\alpha$ $(\alpha=1,2)$.
For general U(1)$_F$ charge assignments on right-handed neutrinos $(q_{N_1}~q_{N_2})$, 
the Dirac- and Majorana-mass matrices of neutrinos are given by
\begin{align}
	(m^\nu_D)_{ i\alpha} \sim v_{\rm EW} 
	\begin{pmatrix}
		\epsilon^{q_{L_1}-q_{N_1}} & \epsilon^{q_{L_1}-q_{N_2}} \\
		\epsilon^{q_{L_2}-q_{N_1}} & \epsilon^{q_{L_2}-q_{N_2}} \\
		\epsilon^{q_{L_3}-q_{N_1}} & \epsilon^{q_{L_3}-q_{N_2}}  
	\end{pmatrix},~~~
	(m^N)_{\alpha\beta}\sim M
	\begin{pmatrix}
		\epsilon^{-2q_{N_1}} & \epsilon^{-q_{N_1}-q_{N_2}} \\
		\epsilon^{-q_{N_1}-q_{N_2}} & \epsilon^{-2q_{N_2}}
	\end{pmatrix}.
\end{align}
According to the seesaw mechanism, after integrating out heavy right-handed neutrinos, we obtain the following light neutrino mass matrix:
\begin{align}
	m^\nu_{ij} =m^{\nu}_D \cdot (m^N)^{-1}\cdot (m^{\nu}_D)^T \sim \frac{v_{\rm EW}^2}{M}
	\begin{pmatrix} 
		\epsilon^{2q_{L_1}} & \epsilon^{q_{L_1}+q_{L_2}} &  \epsilon^{q_{L_1}+q_{L_3}}  \\
		 \epsilon^{q_{L_1}+q_{L_2}}  &  \epsilon^{2q_{L_2}}  &  \epsilon^{q_{L_2}+q_{L_3}}  \\
		 \epsilon^{q_{L_1}+q_{L_3}}  &  \epsilon^{q_{L_2}+q_{L_3}}  &  \epsilon^{2q_{L_3}} 
	\end{pmatrix}.  \label{mnu}
\end{align}
It is independent of the charges of right-handed neutrinos.
Note that since the matrix $m^N$ is rank 2, $m^{\nu}_{ij}$ must contain one zero eigenvalue.
It is diagonalized as
\begin{align}
	m_{ij}^\nu &\sim U^\nu {\rm diag}(m^\nu) (U^{\nu})^T
        \nonumber \\
	&\sim \frac{v_{\rm EW}^2}{M}
	\begin{pmatrix}
		1 &\epsilon^{q_{L_1}-q_{L_2}} & \epsilon^{q_{L_1}-q_{L_3}} \\
		\epsilon^{q_{L_1}-q_{L_2}} & 1 & \epsilon^{q_{L_2}-q_{L_3}} \\
		\epsilon^{q_{L_1}-q_{L_3}} & \epsilon^{q_{L_2}-q_{L_3}} & 1 
	\end{pmatrix}
	\begin{pmatrix}
		 0 & 0 & 0\\
		0 &  \epsilon^{2q_{L_2}} & 0\\
		0 & 0 & \epsilon^{2q_{L_3}}
	\end{pmatrix}
	\begin{pmatrix}
		1 &\epsilon^{q_{L_1}-q_{L_2}} & \epsilon^{q_{L_1}-q_{L_3}} \\
		\epsilon^{q_{L_1}-q_{L_2}} & 1 & \epsilon^{q_{L_2}-q_{L_3}} \\
		\epsilon^{q_{L_1}-q_{L_3}} & \epsilon^{q_{L_2}-q_{L_3}} & 1 
	\end{pmatrix}.
\end{align}
The MNS matrix is given by
\begin{align}
	U_{\rm MNS} = U^\nu V^{l\dagger} \sim \begin{pmatrix}
		1 &\epsilon^{q_{L_1}-q_{L_2}} & \epsilon^{q_{L_1}-q_{L_3}} \\
		\epsilon^{q_{L_1}-q_{L_2}} & 1 & \epsilon^{q_{L_2}-q_{L_3}} \\
		\epsilon^{q_{L_1}-q_{L_3}} & \epsilon^{q_{L_2}-q_{L_3}} & 1 
	\end{pmatrix}.  \label{MNS}
\end{align}
Therefore, the large $\nu_\mu-\nu_\tau$ mixing is obtained for $q_{L_2}=q_{L_3}$.
A reasonable choice to reproduce the observed MNS matrix is thus
\begin{align}
	\begin{pmatrix}
		q_{L_1} & q_{L_2} & q_{L_3}
	\end{pmatrix}
	=
	\begin{pmatrix}
		q_{L_3}+1 & q_{L_3} & q_{L_3}
	\end{pmatrix}.
\end{align}
For $M \sim 10^{14}\text{--}10^{15}$\,GeV as a representative value as described in the main text, 
the observed neutrino mass differences are consistent with $q_{L_3}=0$. This is the one given in (\ref{L_charge}).
For $M \sim 10^{16}\text{--}10^{17}$\,GeV, a slightly small Yukawa coupling $y^N\sim O(0.01)$ is required.
Note that if $q_{L_3}$ takes a half-integer value, all the lepton and right-handed neutrino charges should also be half-integer.

Next, let us consider the case of three right-handed neutrinos: $N_\alpha$ $(\alpha=1$--$3)$.
For general U(1)$_F$ charge assignments on right-handed neutrinos $(q_{N_1}~q_{N_2}~q_{N_3})$, 
the Dirac- and Majorana-mass matrices of neutrinos are given by
\begin{align}
	&(m^\nu_D)_{ i\alpha} \sim v_{\rm EW} 
	\begin{pmatrix}
		\epsilon^{q_{L_1}-q_{N_1}} & \epsilon^{q_{L_1}-q_{N_2}} &  \epsilon^{q_{L_1}-q_{N_3}} \\
		\epsilon^{q_{L_2}-q_{N_1}} & \epsilon^{q_{L_2}-q_{N_2}} &  \epsilon^{q_{L_2}-q_{N_3}} \\
		\epsilon^{q_{L_3}-q_{N_1}} & \epsilon^{q_{L_3}-q_{N_2}} &  \epsilon^{q_{L_3}-q_{N_3}} 
	\end{pmatrix},\\
	&(m^N)_{\alpha\beta}\sim M
	\begin{pmatrix}
		\epsilon^{-2q_{N_1}} & \epsilon^{-q_{N_1}-q_{N_2}} & \epsilon^{-q_{N_1}-q_{N_3}}  \\
		\epsilon^{-q_{N_1}-q_{N_2}} & \epsilon^{-2q_{N_2}} & \epsilon^{-q_{N_2}-q_{N_3}} \\
		\epsilon^{-q_{N_1}-q_{N_3}} & \epsilon^{-q_{N_2}-q_{N_3}} & \epsilon^{-2q_{N_3}}
	\end{pmatrix}.
\end{align}
The resulting structure of the light neutrino mass matrix after integrating out the heavy right-handed neutrinos is the same as (\ref{mnu}).
The MNS matrix is also the same as (\ref{MNS}). Only the difference is that there is no zero mass eigenvalues in the light neutrino mass matrix:
\begin{align}
	m_{ij}^\nu &\sim U^\nu {\rm diag}(m^\nu) (U^{\nu})^T
        \nonumber \\
	&\sim \frac{v_{\rm EW}^2}{M}
	\begin{pmatrix}
		1 &\epsilon^{q_{L_1}-q_{L_2}} & \epsilon^{q_{L_1}-q_{L_3}} \\
		\epsilon^{q_{L_1}-q_{L_2}} & 1 & \epsilon^{q_{L_2}-q_{L_3}} \\
		\epsilon^{q_{L_1}-q_{L_3}} & \epsilon^{q_{L_2}-q_{L_3}} & 1 
	\end{pmatrix}
	\begin{pmatrix}
		 \epsilon^{2q_{L_1}} & 0 & 0\\
		0 &  \epsilon^{2q_{L_2}} & 0\\
		0 & 0 & \epsilon^{2q_{L_3}}
	\end{pmatrix}
	\begin{pmatrix}
		1 &\epsilon^{q_{L_1}-q_{L_2}} & \epsilon^{q_{L_1}-q_{L_3}} \\
		\epsilon^{q_{L_1}-q_{L_2}} & 1 & \epsilon^{q_{L_2}-q_{L_3}} \\
		\epsilon^{q_{L_1}-q_{L_3}} & \epsilon^{q_{L_2}-q_{L_3}} & 1 
	\end{pmatrix}.
\end{align}



\end{document}